\documentclass[10pt,journal]{IEEEtranTCOM}

\usepackage[english]{babel}
\usepackage[usenames]{color}
\usepackage[cp1250]{inputenc}
\usepackage{amsfonts}
\usepackage{amsthm}
\usepackage{graphicx}
\usepackage{epsfig}
\usepackage{mathrsfs}
\usepackage{amsmath}
\usepackage{algorithm}
\usepackage{algorithmic}
\usepackage{hyperref}

\theoremstyle{plain}

\begin{document}
\newcommand{\bea}{\begin{eqnarray}}
\newcommand{\eea}{\end{eqnarray}}
\newcommand{\be}{\begin{equation}}
\newcommand{\ee}{\end{equation}}
\newcommand{\beas}{\begin{eqnarray*}}
\newcommand{\eeas}{\end{eqnarray*}}
\newcommand{\bs}{\backslash}
\newcommand{\bc}{\begin{center}}
\newcommand{\ec}{\end{center}}
\def\SC {\mathscr{C}}

\title{Parametric context adaptive Laplace distribution\\for multimedia compression}
\author{\IEEEauthorblockN{Jarek Duda}\\
\IEEEauthorblockA{Jagiellonian University,
Golebia 24, 31-007 Krakow, Poland,
Email: \emph{dudajar@gmail.com}}}
\maketitle

\begin{abstract}
Data compression often subtracts prediction and encodes the difference (residue) e.g. assuming Laplace distribution, for example for images, videos, audio, or numerical data. Its performance is strongly dependent on the  proper choice of width (scale parameter) of this parametric distribution, can be improved if optimizing it based on local situation like context. For example in popular LOCO-I~\cite{loco} (JPEG-LS) lossless image compressor there is used 3 dimensional context quantized into 365 discrete possibilities treated independently. This article discusses inexpensive approaches for exploiting their dependencies with autoregressive ARCH-like context dependent models for parameters of parametric distribution for residue, also evolving in time for adaptive case. For example tested such 4 or 11 parameter models turned out to provide similar performance as 365 parameter LOCO-I model for 48 tested images. Beside smaller headers, such reduction of number of parameters can lead to better generalization. In contrast to context quantization approaches, parameterized models also allow to directly use higher dimensional contexts, for example using information from all 3 color channels, further pixels, some additional region classifiers, or from interleaving multi-scale scanning - for which there is proposed Haar upscale scan combining advantages of Haar wavelets with possibility of scanning exploiting local contexts.
\end{abstract}
\section{Introduction}
Many types of data statistically agree with specific parametric distributions, like Gaussian distribution through the law of large numbers, or Laplace distribution popular in data compression as it agrees with statistics of errors from prediction (residues). Their parameters can often be inexpensively estimated, and storing them in a header is much less expensive than e.g. entire probability distribution on some quantized set of represented values. Parametric distributions smoothen between discretized possibilities, generalizing statistical trends emerging in a given type of data.

However, for example due to randomness alone, statistics of real data  usually has some distortion from such idealization. Directly storing counted frequencies can exploit this difference, gaining asymptotically Kullback-Leibler divergence bits/value - at cost of larger header. Data compressors need to optimize this minimum description length~\cite{MDL} tradeoff between model size and entropy it leads to.

In practice, instead of a single e.g. Laplace distribution to encode residues (errors of predictions) for the entire image, we would like to make its parameters dependent on local situation - through context dependence like in Markov modelling, or adaptivity as for non-stationary time series.

\begin{figure}[t!]
    \centering
        \includegraphics{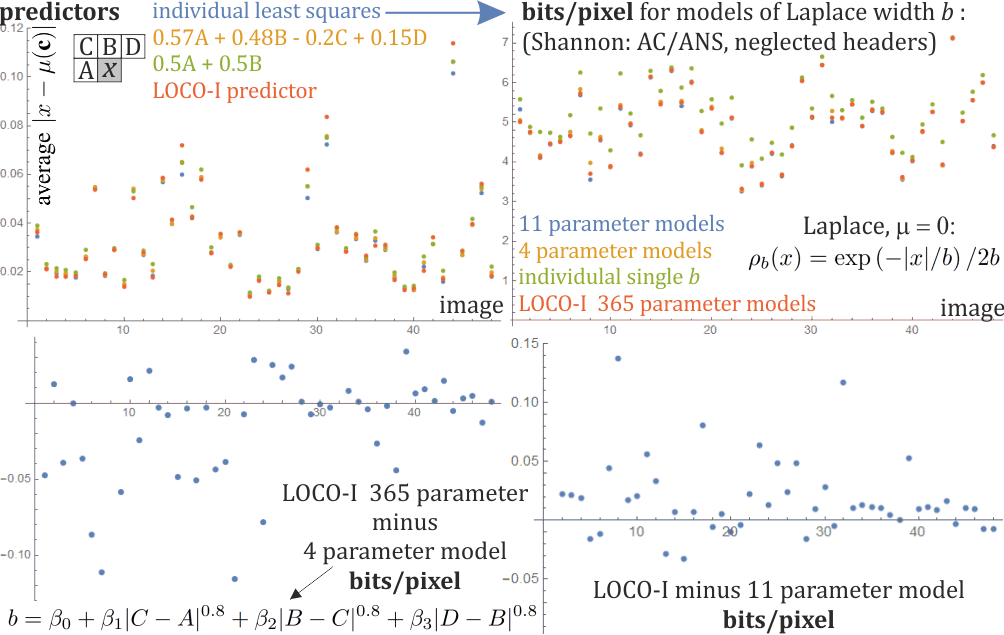}
        \caption{Comparison of some discussed models for 48 grayscale 8 bit 512x512 images presented in Fig. \ref{img}. Top left: first we need to predict pixel value based on the current context: already decoded 4 neighboring pixels $c=(A,B,C,D)$. This predicted $\mu(c)$ is used as the center of Laplace distribution, which is estimated as median: minimizes $l^1$ distance. Hence, presented evaluation uses average $|x-\mu(c)|$  for 4 approaches: LOCO-I predictor (red), simple average (green), least squares parameters for combined images (orange), and least squares parameters chosen individually for each image (blue) - the last one gives the lowest residues so it is used further. Top right: bits/pixel for encoding its residues $(r=x-\mu(c))$ using centered $(\mu=0)$ Laplace distribution of width (scale parameter) $b$ modeled in various ways. Red: LOCO-I model with 365 parameters corresponding to quantized context: $(|C-A|, |B-C|, |D-B|)$. Green: single $b$ chosen individually (MLE) for each image. Orange: discussed here 4 parameter model, written at the bottom left, blue: discussed later 11 parameter model. Bottom: differences of these values for the two models. The evaluation assumes accurate entropy coding (AC/ANS) and neglects headers - including them would worsen especially LOCO-I evaluation if storing all 365 parameters. }
        \label{comp}
\end{figure}

The possibility to directly store all values fades away when increasing dimension of the model - both due to size growing exponentially with dimension, but also underrepresentation. Going to higher dimensions requires finding and exploiting some general behaviour, for example through parametrizations, as in examples presented in Fig. \ref{comp}.\\

LOCO-I\cite{loco} mixes both philosophies: uses parametric probability distributions, which scale parameter (width of Laplace distribution) depends on 3 dimensional context quantized into 365 possibilities treated independently - neglecting their dependencies. Such approach is useful for low dimensional contexts, however, it becomes impractical if wanting to use higher dimensional context, e.g.: using information from all 3 color channels, further pixels than the nearest neighbors, or from some region classifiers to gradually transit between e.g. models for smooth regions like sky, to complex textures like treetop. Finally contexts of much higher dimension appear in multiscale interlaced scanning like in FLIF~\cite{flif} compressor: progressively improving quality, rather only parametric models can directly work on its high dimensional contexts.

This article discusses such parametric-parametric models: choose parameters of e.g. Laplace distribution as a parametric function of the context, like through a linear combination, or generally e.g. neural networks. Its example are ARMA-ARCH~\cite{arma} models popular in economics: choosing squared width of Gaussian distribution as a linear combination of recent squared residues, e.g.  $\sigma^2_t=\beta_0+\beta_1 \epsilon^2_{t-1}$.

These parameters can be universal e.g. default for various types of classified regions, or optimized individually by compressor and stored in the header. For the latter purpose we will focus on least squares estimation due to its low cost. Presented test results are for such estimation, a costly additional optimization might slightly improve performance.

While we will mostly focus on such static models: assuming constant joint distribution of (value, context), mentioned alternative are adaptive models: assuming non-stationary time series, evolving joint distribution. It requires additional cost to update parameters of the model, for example performing likelihood  optimization step while processing each value. It has two advantages: can learn model from already decoded data even without header, and can flexibly adapt to local behavior e.g. of an image. Appendix discusses second order approaches for such online optimization.

In literature there are also considered much more costly models, like using neural networks for predicting probability distribution of succeeding pixels~(\cite{sub1,sub2}). In the discussed philosophy, instead of directly predicting probability of each discrete value e.g. with softmax, we can use such neural networks to directly predict context dependent parameters of some parametric distribution for the new pixel. Such simplification should allow to use much smaller neural networks, bringing it closer to practical application in data compression.

 \begin{figure}[t!]
    \centering
        \includegraphics{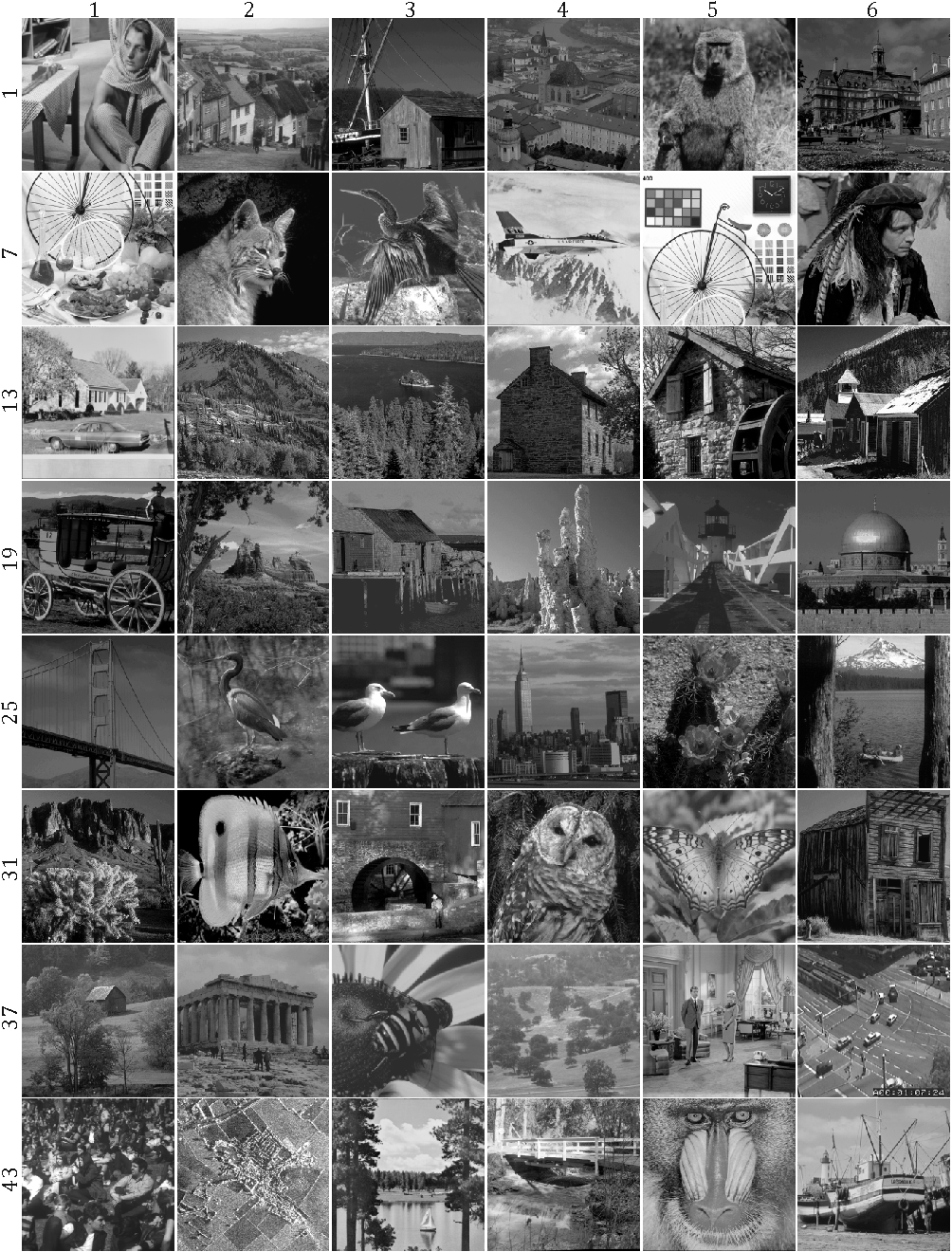}
        \caption{Dataset of 48 grayscale 8 bit 512x512 images used in tests. Source: \url{http://decsai.ugr.es/cvg/CG/base.htm} .}
        \label{img}
\end{figure}

\section{Parametric-parametric distributions}
We would like to model conditional probability distribution $\textrm{Pr}(x|c)$ of the new value $x\in \mathbb{R}$, based on some local $d$-dimensional context $c=(c_1,\ldots,c_d)\in\mathcal{C}\subset\mathbb{R}^d$, in practice bounded e.g. to a cube like $\mathcal{C}=[0,1]^d$ here. In LOCO-I image compressor this context are 4 neighboring already decoded pixels ($c=(A,B,C,D)$ as in Fig. \ref{comp}). Both value and context are rather discrete through some quantization, but it is useful to model them as real values - especially wanting to exploit continuity of their behavior.

Modelling general continuous conditional distributions is a difficult task - requires techniques like quantile regression~\cite{quantile}. or hierarchical correlation reconstruction~\cite{hcr,cred}. However, the situation becomes much simpler if focusing on simple parametric distributions for the predicted distribution. Another standard simplification is separately modelling the center of the distribution with predictor $\mu(c)$, and the remaining parameter(s) $\theta(c)$ of centered distribution for $r=x-\mu(c)$ residue, usually single scale parameter defining width:
\be r=x-\mu(c)\quad \textrm{residue from}\quad \rho_{\theta(c)}\quad \textrm{density}\label{split}\ee

We will mainly focus on standard for such applications Laplace distribution and modeling its width parameter $b$:
\be \rho_{\mu b}(x)=\frac{1}{2b}\exp\left(-\frac{|x-\mu|}{b} \right)\qquad \rho_{b}(r)=\frac{1}{2b}\exp\left(-\frac{|r|}{b}\right) \ee
which MLE parameters for $(x^1,\ldots,x^n)$ sample are:
\be \mu=\textrm{median of }\{x^i\}\qquad b=\frac{1}{n}\sum_{i=1}^n |x^i-\mu|\ee
LOCO-I has a fixed specialized predictor. Then chooses width parameter $\theta(c)\equiv b(c)$ as locally constant inside 365 regions for quantized $|C-A|,|B-C|,|D-B|$ context, each into 9 ranges of nearly equal population. This way we can perform estimation independently for each region, and finally e.g. store in the header the 365 parameters.

Quantization of context neglects dependencies between these regions and can be practical rather only for low dimensional contexts - both due to the number of possibilities growing exponentially with dimension, but also underrepresentation of many such contexts. To resolve it, we will focus here on parameterized models for these parameters:

$$ \mu(c)\equiv\mu_\alpha(c)\quad\textrm{for  } \alpha\in \mathbb{R}^{d_\alpha} \qquad\textrm{predictor}$$
$$ \theta(c)\equiv\theta_\beta(c)\quad\textrm{for  }\beta\in\mathbb{R}^{d_\beta} \qquad\textrm{e.g. scale parameter}$$
Choosing $\mu_\alpha(c)$ and $\theta_\beta(c)$ family of functions optimized for a given type of problems is a difficult question. Like ARCH, unlike LOCO-I, we will focus on using linear combinations of some chosen $f,g$ functions:
\be \mu_\alpha(c)=\alpha_1 f_1(c)+\alpha_2 f_2(c)+\ldots+\alpha_{d_\alpha}\,f_{d_\alpha}(c) \label{pred}\ee
\be \theta_{\beta}(c)=\beta_1 g_1(c)+\beta_2 g_2(c)+\ldots +\beta_{d_\beta}\, g_{d_\beta}(c) \label{be}\ee
The latter might need additional e.g. $\max(\theta,0.001)$ if positive values are required and some of $\beta$ are negative. We can alternatively use more sophisticated nonlinear models like neural networks.
\subsection{Context dependence}
Choosing some $\mu_\alpha(c)$ and $\theta_\beta(c)$ family of functions, we can optimize $\alpha, \beta$ (or e.g. neural network parameters) for given $(x^1,\ldots,x^n)$ values and $(c^1,\ldots,c^n)$ contexts, for example maximizing likelihood (MLE):
\be(\alpha,\beta)=\textrm{argmin}_{\alpha,\beta} \sum_{i=1}^n \log\left(\rho_{\theta_{\beta}(c^i)}(x^i-\mu_\alpha(c^i))\right) \label{comb}\ee
To simplify this optimization at cost of suboptimality, we can split it into predictor and the remaining as in Fig. \ref{split}.

This way we can first optimize parameters of predictor e.g. using some distance $d$:
\be\alpha=\textrm{argmin}_{\alpha} \sum_{i=1}^n d(x^i,\mu_{\alpha}(c^i)) \ee
for example using $d(x,y)=(x-y)^2$ least squares distance we are looking for predictor of expected value - appropriate e.g. for Gaussian distribution (or polynomial coefficients in \cite{cred}). For Laplace distribution it is more appropriate to use $d(x,y)=|x-y|$ for predictor of median. However, unless heavy tails case, optimization of both gives nearly the same predictor, so it is safe to use least squares optimization which is computationally less expensive.

Having optimized predictor, we can calculate residues $r^i=x^i-\mu_\alpha(c^i)$ and separately optimize $\beta$ using them. Especially for scale parameter, MLE estimator is often average over some simple function of values, for example $b=$ average $|r|$ for Laplace distribution $(\theta\equiv b)$, $\sigma^2$ = average $r^2$ for Gaussian distribution $(\theta\equiv \sigma^2)$, or generally average $|r|^\kappa$ for exponential power distribution $(\theta\equiv b^\kappa)$. Average is estimator of expected value, what allows for practical optimization of $\beta$ using least squares (analogously e.g. for neural networks):
\be \beta=\textrm{argmin}_\beta \sum_{i=1}^n \left(|r^i|-\theta_\beta (c^i)\right)^2\ \textrm{for Laplace:}\ \theta\equiv b \label{est}\ee
$$ \beta=\textrm{argmin}_\beta \sum_{i=1}^n \left((r^i)^2-\theta_\beta (c^i)\right)^2\ \textrm{for Gaussian:}\ \theta\equiv\sigma^2$$

Such parameters can be optimized for a dataset, for example for different regions using some segmentation, and then used as default. Alternatively, compressor can optimize them individually e.g. for a given image and store parameters in the header.

\subsection{Adaptivity}
Instead of storing model parameters in the header, alternative approach is starting from some default parameters and adapting them based on the processed data, also for better agreement with varying local statistics e.g. of an image. Such adaptation brings additional cost, dependence on local situation can be alternatively realized by using some region classifier/segmentation and separate models for each class, or using outcome of such local classifier as additional context - choosing the best tradeoffs is a difficult question.

For adaptation we can treat the upper index as time and use time dependent parameters starting from some e.g. default initial choice for $t=0$. For example without context dependence, we could just replace average with exponential moving average for Laplace distribution and some $\eta,\nu\in(0,1)$ learning rates:
$$ \mu^{t+1}=\nu\mu^t +(1-\nu)x^t\qquad b^{t+1}=\eta b^t +(1-\eta)|x^t-\mu^t|$$

Generally we could use for example gradient descent while processing each value to optimize parameters toward local statistics for combined $(\alpha,\beta)$ using (\ref{comb}), or in split form:
$$r^t=x^t-\mu_{\alpha^t}(c^t) \qquad \textrm{residue from}\qquad \rho_{\theta_{\beta^t}}\quad\textrm{density}$$
$$\alpha^{t+1}=\alpha^t-\eta_\alpha \frac{\partial d(x^t,\mu_{\alpha} (c^t))}{\partial \alpha}(\alpha^t)  $$
\be \beta^{t+1}=\beta^t+\eta_\beta \frac{\partial \log(\rho_{\theta_\beta (c^t)}(r^t))}{\partial \beta} (\beta^t) \ee
where $d$ is distance as previously. For $\beta$ the above gradient ascend optimizes likelihood, $\eta_\alpha, \eta_\beta$ define adaptation rate.

Using first order method is not sufficient for a proper choice of step size, suggesting to use also second derivative and Newton's method (e.g. $\forall_i\ \theta_i^{t+1}=\theta_i^t-\partial_i f(\theta^t)/\partial_{ii} f(\theta^t)$) - the Appendix discusses such general approaches.


\subsection{Exponential power distribution}
Data compression usually focuses on Laplace distribution, but real data might have a bit different statistics, especially heavier tails. It might be worth to consider more general families, especially exponential power distribution~\cite{exp}:
\be \rho_{\kappa\mu b}(x)=\frac{\kappa^{-1/\kappa}}{2\,b\, \Gamma(1+1/\kappa)} e^{-\frac{1}{\kappa}\left(\frac{|x-\mu|}{b}\right)^\kappa} \ee
It covers both Laplace $(\kappa=1)$ and Gaussian $(\kappa=2, b\equiv\sigma)$ distribution. Estimating $\kappa$ is costly, but we can fix it based on a large dataset and e.g. segment type. Then estimation of $\mu,b$ is analogous, also for context dependence like in \ref{est}:
$$\mu=\textrm{argmin}_\mu \sum_{i=1}^n |x^i-\mu|^\kappa\qquad b=\left(\frac{1}{n}\sum_{i=1}^n |x^i-\mu|^\kappa\right)^{1/\kappa}$$
\be\beta=\textrm{argmin}_\beta \sum_{i=1}^n \left(|r^i|^\kappa-\theta_\beta (c^i)\right)^2\quad\textrm{for}\quad\theta\equiv b^\kappa \ee
Here is a simple example of its adaptive estimation for $\eta,\nu\in(0,1)$ learning rates:
$$ \mu^{t+1}=\nu\,\mu^t +(1-\nu)\,x^t $$
\be \theta^{t+1}=\eta\, \theta^t +(1-\eta)\,|x^t-\mu^t|^\kappa\qquad\textrm{for }\theta\equiv b^\kappa\label{aEPD}\ee
In data compression we can have prepared entropy coding tables for such fixed $\kappa$ and some optimized discretized set of scale parameter $b$.

\subsection{Adaptive least-squares linear regression$^*$} 
\begin{small}$^*$This subsection expands adaptivity to linear regression - for completeness and to connect some concepts, however, it might be too costly for data compression and is not used further (yet).\end{small}

Above (\ref{aEPD}) formula for $b$ can be seen as obtained from online adaptive ML estimation: instead of standard "static" estimation of constant parameters based on the entire sample, we perform ML estimation separately for every moment in time - using only its past information, weakening influence of old values e.g. with exponential moving average. This way we optimize parameters separately for every time, instead of standard: finding a single compromise for all of them.

Specifically, we get (\ref{aEPD}) formula for $b$ if maximizing \be l^T=\sum_{t<T} \eta^{T-t} \lg(\rho^t(x^t))\quad \textrm{weighted likelihood}\ee in time $T$, for fixed $\kappa$ and $\mu$, separately for each time $T$.

We could perform such optimization using some gradient ascend, however, it would be beneficial to have a direct formula like for scale parameter $b$ of exponential power distribution. Getting such useful direct formulas is relatively difficult, above is for MLE, adaptivity for polynomial as model of density in \cite{hcr} can be seen as using MSE instead.\\

Let us now discuss another basic MSE adaptivity situation with direct evolution formulas: least-squares linear regression e.g. for \cite{cred} approach, starting with generalization of least-squares linear regression formulas to weighted case.

For time series of values $(x^t)_t$ and their contexts $(M_{ti})_{ti}$, in time $T$ we would like to find parameters $\beta\equiv \beta^T$ 
\be\beta^T=\textrm{argmin}_\beta\ \sum_{t<T} \eta^{T-t} ((M\beta)_t-x^t)^2\label{alsr}\ee
 using only $t<T$ values, with exponentially weakening weights $w_t=\eta^{T-t}$.

Here is derivation of parameters for general weighted least-square linear regression with weights $(w_t)_t$:

$$0=\partial_{\beta_j}\sum_{t} w_t \left(\sum_i M_{ti}\beta_i -x^t\right)^2=$$
$$=2\sum_{t} w_t \left(M_{tj}\left(\sum_i M_{ti}\beta_i -x^t     \right)\right)$$
Leading to general formula for weighted linear regression:
\be \beta=(M^\dagger \textrm{diag}(w) M)^{-1} M^\dagger \textrm{diag}(w)  x\label{wlr}\ee
where $\textrm{diag}(w)_{ij}=\delta_{ij}w_i$ is diagonal matrix, $\dagger$ denotes transposition.\\

For adaptive linear regression, in time $T$ we can use only information from times $t<T$, and it is convenient to use exponential moving average weights: $w_t=\eta^{T-t}$. We could just insert it to (\ref{wlr}) getting $\beta^T$ for time $T$.

Let us try to find a recurrence relation for more efficient calculation. Denoting $y^T$ as $M^\dagger \textrm{diag}(w)x $ in time $T$, we get:
\be y^{T+1}= \eta\left( y^T +x^T M_{T\bullet}\right)\label{rec1}\ee
where $M_{T\bullet}$ denotes vector: $(M_{T\bullet})_i=M_{Ti}$. Analogously denoting $\mathcal{M}^T$ as $M^\dagger \textrm{diag}(w) M$ in time $T$, we get:
\be \mathcal{M}^{T+1}=\eta\left(\mathcal{M}^T+ (M_{T\bullet}) (M_{T\bullet})^\dagger \right)\label{rec2}\ee
Recurrences (\ref{rec1}), (\ref{rec2}) lead to parameters (\ref{wlr}) for time $T$:
\be \beta^T = (\mathcal{M}^{T})^{-1} y^T\ee
We can start e.g. with zero $y^0$ and $\mathcal{M}^0$, then there is needed a warmup: some number of steps (at least the number of indexes to make $\mathcal{M}$ invertible) when we update $y$ and $\mathcal{M}$, but not use linear regression.

To avoid matrix inversion, it might be worth to consider recurrence for $(\mathcal{M}^{T})^{-1}$ instead:
$$(\mathcal{M}^{T+1})^{-1}=\eta^{-1}\left(\textbf{1}+(M_{T\cdot}) (M_{T\cdot})^\dagger (\mathcal{M}^{T})^{-1}\right)^{-1} (\mathcal{M}^{T})^{-1}$$
$$\beta^{T+1}=(\mathcal{M}^{T+1})^{-1} y^{T+1}=$$
$$\left(\textbf{1}+(M_{T\cdot}) (M_{T\cdot})^\dagger (\mathcal{M}^{T})^{-1}\right)^{-1}\left(\beta^T+x^T(\mathcal{M}^{T})^{-1} M_{T\bullet}  \right)$$
e.g. using some $1/(1+z)=1-z+\ldots$ expansion as approximation.

\section{Practical Laplace example and experiments}
Let us now focus on LOCO-I lossless image compression setting: context are 4 already decoded neighboring pixels: $c=(A,B,C,D)$  on correspondingly (left, up, left-up, right-up) positions as in diagram in Fig. \ref{comp}.
\subsection{Predictor $\mu(c)$}
LOCO-I uses a fixed predictor $(c=(A,B,C,D))$:
\be \mu(c)=\left\{\begin{array}{ll}
\min(A,B)\quad \,\,\textrm{if }C\geq \max(A,B)\\
\max(A,B)\quad \,\textrm{if }C\leq \min(A,B)\\
A+B-C\quad \textrm{otherwise}
\end{array}\right.\ee
Simpler popular choices are e.g. $(A+B)/2$ or $A+B-C$. A standard way for designing such predictors is polynomial interpolation, e.g. in Lorenzo predictor~\cite{lorenzo}: fitting some polynomial to the known values and calculating its value in the predicted position, getting a linear combination.

We can also directly optimize it for a dataset. For example least squares optimization using combined 48 images (Fig. \ref{img}) gives (rounded to 2 digits, weights sum to 1):
$$\mu(c)=0.57A+0.48B-0.2C+0.15D$$
Alternatively, we can optimize these weights individually for each image by compressor and store in the header - Fig. \ref{comp} contains comparison for various approaches using $l^1$ distance as we would like to estimate median for Laplace distribution. Such individual least squares optimization turns out always superior there (blue points), LOCO-I predictor for some images is much worse than the remaining.

Tested inexpensive least squares optimizer uses directly the $d_\alpha=4$ functions: $f_1(c)=A,f_2(c)=B,f_3(c)=C,f_4(c)=D$ in \ref{pred} notation. We build $n\times d_\alpha$ matrix $P$ from them: $P_{ij}=f_j(c^i)$, and $x=(x^1,\ldots,x^n)$ vector. Then the optimal parameters are obtained using pseudo-inverse (as derived (\ref{wlr}) for equal weights $w$):

\be\alpha=\textrm{argmin}_\alpha \|P\alpha-x\|_2^2=(P^\dagger P)^{-1} P^\dagger x\ee
For further tests there were used residues from individual least squares optimization for each image: $r=x-P\alpha$.
\subsection{Context dependent scale parameter $b(c)$}
Having the residues, LOCO-I would divide $|C-A|, |B-C|, |D-B|$ into 9 ranges each, having nearly equal population. Including symmetry it leads to division into $(9^3+1)/2=365$ contexts. For each of them we independently estimate scale parameter $b$ of Laplace distribution.

Here we would like to model $b$ as a linear combination (\ref{be}) of some functions $(g_j(c))_{j=1..d_\beta}$ of the context. The choice of these functions is difficult and essentially affects compression ratios. They should contain "1" for the intercept term. Then, in analogy to LOCO-I, the considered 4 parameter model uses the following linear combination (for convenience enumerated from 0):
\be b(c)=\beta_0 +\beta_1 |C-A|^{0.8}+\beta_2 |B-C|^{0.8}+\beta_3 |D-B|^{0.8}\label{eq}\ee

\begin{figure}[t!]
    \centering
        \includegraphics{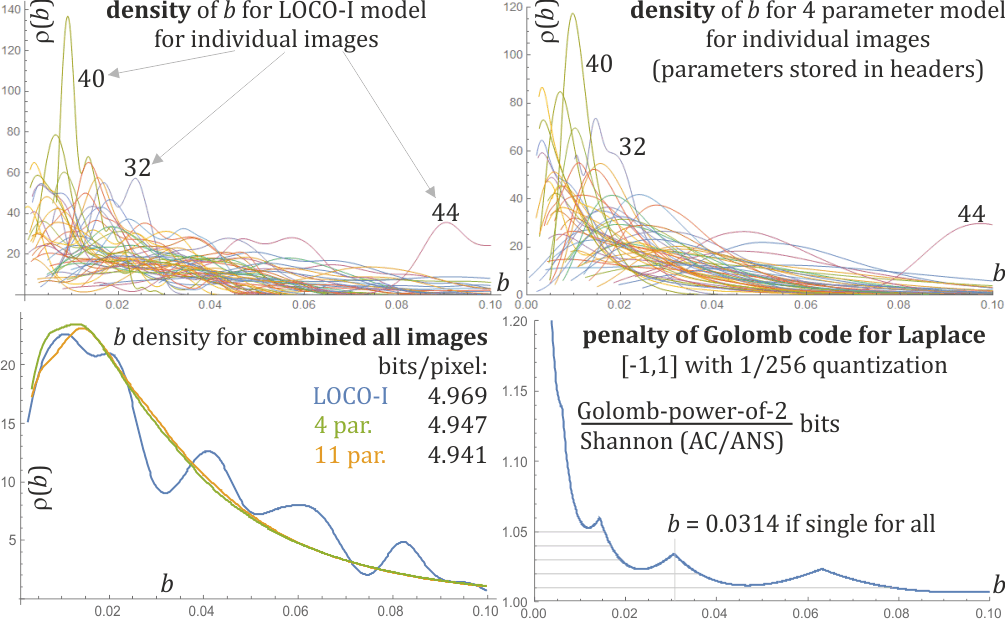}
        \caption{Top: probability density of $b$ parameters for all images, LOCO-I and discussed 4 parameter model, assuming the models are estimated and stored individually for each image. Three most characteristic images are marked as their numbers. Bottom left: such densities if combining all images into one - while huge LOCO-I number of parameters can usually learn better individual images than 4 parameter model, it has worse generalization - is inferior when combining different types of patterns. Bottom right: penalty of using power-of-2 Golomb coding for various $b$ parameters. We can get $\approx 2\%$ improvement if switching to arithmetic coding or asymmetric numeral systems, however, especially for LOCO-I it would require larger headers due to needed better precision of $b$. }
        \label{bpar}
\end{figure}

There is a freedom of choosing above power and empirically $\approx 0.8$ has turned out to provide the best likelihood/compression ratio - corresponds well to linear behavior of $b$. This choice leads to all the coefficients $\beta$ turn out positive in experiments - we have some initial $\beta_0$ width, growing with increased gradients in the neighboring pixels. Hence there is no possibility of getting negative $b$ this way, which would make no sense.

Having chosen such e.g. $d_\beta=4$ functions, we build $n\times d_\beta$ matrix from them $S_{ij}=g_j(c^i)$, and residue vector $|r|=(|r^1|,\ldots,|r^n|)$. Then we can use least squares optimization:
\be\beta=\textrm{argmin}_\beta \|S\beta-|r|\|_2^2=(S^\dagger S)^{-1} S^\dagger |r| \label{eqq}\ee

Figure \ref{bpar} contains comparison of density of predicted scale parameters $b$ for individual images (top) for LOCO-I approach and the above 4 parameter model - the latter is smoother as we could expect, but generally they have similar behavior. Bottom left of this figure contains comparison for combining all images, and compression ratios showing better generalization of these low parameter models.

The second considered: $d_\beta=11$ parameter model extends above basis by the following arbitrarily chosen 7 functions: symmetric describing intensity of neighboring pixels, and evaluating the second derivative:
$$ (A-0.5)^4,(B-0.5)^4,(C-0.5)^4,(D-0.5)^4$$
\be |C-2B+D|^{0.1},|A-2C+B|^{0.1} \ee
where again powers were chosen empirically to get the best likelihood/compression ratio. In contrast to 4 parameter model, this time we get also negative $\beta$ coefficients, leading to negative predicted $b$. To prevent that, there was finally used $\max(b,0.001)$ width of Laplace distribution.\\

The used functions  were chosen arbitrarily by manual optimization, some wider systematic search should improve performance. For example in practical implementations above power functions would be rather put into tables, what allows to use much more complex functions, like given by stored values on some quantized set of arguments. It would allow to carefully optimize such tabled functions based on a large set of images.

The above was for Laplace distribution. For more general exponential power distribution, there should be used $|r|^\kappa$ in (\ref{eqq}) instead of $|r|$, and the prediction $S\beta$ like (\ref{eq}) gives $b^\kappa$.
\subsection{Entropy coding, penalty of Golomb coding}
Laplace distribution is continuous, to encode values from it we need to quantize it to approximately geometric distribution, which values are transformed into bits using some entropy coding.

LOCO-I uses power-of-2 Golomb coding: instead of real $b$ coefficient, it optimizes $M=2^m$ parameter, then $x$ is stored as $\lfloor x/M\rfloor$ using unary coding, and $\mod(x,M)$ is stored directly as bits. This way it requires $2\lfloor x/M\rfloor+1+m$ bits to store unsigned $x$. Signed values are stored as position in $0,1,-1,2,-2,\ldots$ order.

Ideally, symbol of probability $p$ carries $\log_2(1/p)$ bits of information, leading to asymptotically Shannon entropy bits/symbol. Optimal parameter power-of-two Golomb coding is worse by a few percents for used here $b$ values as shown in Fig. \ref{bpar}. One reason is this sparse $M=2^m$ quantization of parameters. More important, especially for small $b$, is most of probability going to $0$ quantized value, what can correspond to lower than 1 bit of informational content. In contrast, prefix codes like Golomb need to use at least 1 bit per symbol.

Replacing power-of-2 Golomb coding with an accurate entropy coder like arithmetic coding (AC) or asymmetric numeral systems (ANS), we can improve compression ratio by $\approx 2\%$. In this case we also need some quantization of $b$ parameter - we can have prepared entropy coding tables for some discredited space of possible parameters.

\begin{figure}[b!]
    \centering
        \includegraphics{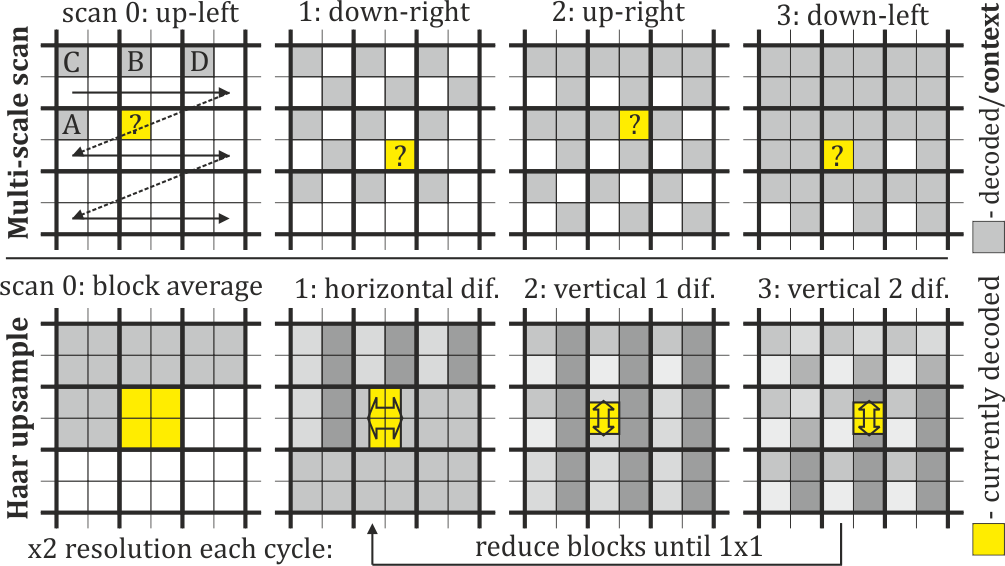}
        \caption{Top: conventional multi-scale interleaved scanning~\cite{sub1} (e.g. FLIF compressor~\cite{flif}): scan over succeeding sub-lattices for progressive decoding, and most importantly: to provide better local context for later decoded pixels. Bottom: proposed Haar upsample scanning which combines advantages of Haar wavelets~\cite{haar} with exploitation of local context dependence. First (scan 0) we decode low resolution image: averages over $2^k\times 2^k$ size blocks, using decoded neighboring block averages as the context. Then in each cycle (scan 1,2,3) we decode the 3 missing values (for grayscale, 9 for RGB) to improve the resolution twice: e.g. horizontal differences in scan 1, then vertical differences in two positions in scan 2 and 3. After $k$ such cycles we reach $1\times 1$ blocks - completely decoded image. The context of already decoded local information is high dimensional, of different type for each scan and level. While it is a problem for LOCO-I like context quantization, parametric models can easily handle it, for example using $\mu^s(c)=\sum_i \alpha^s_i c_i$ predictor, where $s$ denotes the type of scan - its parameters $\alpha$ can be inexpensively e.g. MSE optimized and stored in the header. Some modification options are e.g. splitting values into higher and lower bits for separate scans~\cite{sub2}, or using fractal-like (tame twindragon) blocks by modifying translation vectors for hexagonal block lattice~\cite{frac}. }
        \label{multi}
\end{figure}

\subsection{Multi-scale interleaving}
In standard scanning line by line we have context only from half of the plane, only guessing what will happen from the decoded side. It can be improved in multi-scale interleaving, showing gains e.g. in FLIF~\cite{flif} compressor, where we can use lower resolution context from all directions due to progressive decoding in multiple scans, like visualized in Fig. \ref{multi}.

However, we can see that context information becomes much more complex here: high dimensional, varying with the scan number. Even reducing it by some arbitrary averaging, it is still rather too large for context quantization approaches like in LOCO-I. Discussed here parametric approaches have no problem with direct use of such high dimensional contexts, modelling parameters as e.g. a linear combination of a chosen family of functions, with parameters chosen e.g. by inexpensive least squares optimization and stored in the header. Alternatively more complex models can be used instead, like neural networks.

This Figure also proposes combination with Haar wavelets for hopefully improved performance - splitting decoding into $k$ cycles, each improving resolution twice, and being composed of a few scans, e.g. 3 for grayscale, or 9 for 3 colors - each providing a single degree of freedom per block for the upscaling. Such decomposition into e.g. 9 scans clearly leaves an opportunity for optimization, starting with the choice of color transformation.

Assuming some scale invariance of images, similar models can be used for different cycles here, for example we can treat the number of cycle (defining scale) as an additional parameter.

\section{Conclusion and further work}
Parametric models allow to successfully exploit trends in behavior, also for context dependence and evolution of parametric distributions. Thanks to generalization, a few parameter model can provide a better performance than treating all possibilities as independent - neglecting dependencies between them. Wanting to exploit higher dimensional contexts, e.g. for 3 colors, further pixels, region classifiers or multi-scale scanning, parametric models become a necessity as the number of discretized possibilities would grow exponentially with dimension.

There were presented and tested very basic possibilities, leaving many improvement opportunities, starting with choice of contexts and functions, or using other parametric distributions like exponential power distribution. Used least squares optimization is inexpensive enough to be used by compressor to individually optimize parameters for each image. For example choosing some general default parameters, we can use better optimizers, like $l^1$ for Laplace median, or generally MLE. These parameters can be alternatively optimized online, e.g. with discussed adaptive linear regression, however, it might be too costly for data compression.

Lossy image compressors have a different situation: coding e.g. DCT transform coefficients, where distribution parameters should be chosen also based on position - which should be included as a part of the context with some properly chosen functions.

As we can see in Fig. \ref{bpar}, there is a large spread of behavior of parameters, using individual models for separate images often gives improvement. It suggests to try to  segment the image into regions of similar behavior, or use a region classifier. Having such segmentation mechanism optimized for a large dataset, with separate models for each segment, they could define default behavior, avoiding the need of separate model estimation and storage. It would be valuable to optimize such segmentation based on used family of models. Alternative approach is using classifiers and treating their evaluation as part of the context, what would additionally allow to continuously interpolate between classes.

Finally, while for low cost reason we were focused on linear models for parameters, better compression ratios at larger computational cost should be achievable using more general models like neural networks. They are considered in literature to directly predict discrete probability distributions for pixel values~(\cite{sub1,sub2}). We could reduce the computational cost if, based on the context, predicting only parameters of parametric distributions instead, then finally discretizing the obtained distribution. For example Laplace distribution for unimodal distributions, e.g. training neural network to minimize sum of $|x-\mu(c)|$ and $(b(c)-|x-\mu(c)|)^2$. For more complex distributions like multimodal, we can e.g. parameterize density as polynomial, and train to minimize sum of squares of differences for coefficients of orthonormal polynomials as in \cite{cred}.

\appendix
The appendix discusses some general approaches for online adaptation of parameters of models to optimize for local behavior of non-stationary time series using updated second order local approximation of the optimized function.

For this purpose we should define a function we want to optimize. This function needs to evolve in time to express local situation we would like to optimize for. Recently observed values bring us information about this local situation - we can for example use log-likelihood for them to estimate evolution of parameters of probability distribution, reducing the weights of the old observations to find the current local behavior. It is convenient to use exponentially weakening weights as in exponential moving average, leading e.g. to (\ref{aEPD}) adaptive estimation.

\subsubsection{Second order online parameter optimization}
Let us choose example of such family of online optimized evolving criterion for $(x^t)$ series of observations - for time $T+1$ as:
\be F^{T+1}(\theta)=\sum_{t\leq T} \eta^{T-t} f(x^t,\theta)=\eta F^{T}(\theta)+f(x^{T},\theta) \label{optfun}\ee
using some coefficient $\eta\in(0,1)$ and $f$ point-wise evaluation e.g. logarithm of density for log-likelihood. In practice this sum is finite - requiring to choose some initial value for above recurrence in exponential moving average.

In machine learning such objective/cost functions $F(\theta)$ are usually minimized, so we can use minus logarithm to optimize log-likelihood, or some its approximation like first few terms of Taylor expansion:
$$f(x,\theta)=-\ln(\rho_\theta (x))\qquad (= \sum_{k=1}^{\infty} \frac{(1-\rho_\theta (x))^k}{k} )$$

Now for online minimization of $F$, a natural assumption is that in time $T+1$ we know $\theta^T$ minimizing $F^T$, and want to find $\theta^{T+1}$ minimizing $F^{T+1}$. To reduce cost, we would like to slowly evolve parameters $(\theta^{T+1}\approx \theta^{T})$, what generally requires a caution: might be suboptimal if $F$ is not convex.

To approximate such preferable step $\theta^{T+1}-\theta^T$, we can use derivatives calculated with recurrence as in \ref{optfun}, e.g.:
$$\partial_{\theta_i} F^{T+1}=\eta\, \partial_{\theta_i} F^{T}+\partial_{\theta_i} f(x^{T},\theta)$$
However, to work on values we would need to fix a point $(\theta^*)$ where such derivatives are taken. We could choose this point as some averaged parameters, and use its perturbed values based on online calculated first two derivatives in this point e.g. using Newton method - multidimensional or separately for each coordinate $i$:
\be \forall_i\quad \theta_i^T=\theta^*_i -\partial_{\theta_i}F^T(\theta^*)\,/\,\partial_{\theta_i\theta_i}F^T(\theta^*)\label{newton}\ee

For more general evolution of parameters we need to be able to shift this point of derivation, for example by updating model in two points simultaneously, using the older one to get the actual parameters (\ref{newton}), and periodically replacing such used point with the new one, and starting building a new model for a recent $\theta^*=\theta^T$ point for derivations.

\subsubsection{Adaptive minimization with online parabola model}
To get a more continuous update of parameters, alternative approach might be treating $(\theta^t, \nabla_\theta f(x^t,\theta)|_{\theta=\theta^t})_{t\leq T}$ as a sequence of (value, noisy gradient) like in stochastic gradient descent. Instead of directly using second derivative, we can see neighboring minimum as where the gradient becomes zero - what can be estimated by finding linear trend of gradients and calculating where this trend crosses zero.

Linear trend of gradients can be calculated in online way by using least squares linear regression with exponentially weakening weights~\cite{sgd}. Let us present it in one-dimensional case e.g. to perform optimization separately for each $\theta_i$ parameter, what seems sufficient for slowly evolving minimum. In multidimensional case it can alternatively be done using Hessian inversion as in Newton's method - which is more costly, but might be slightly better.

So let us focus on optimization for 1D parameter $\theta\in\mathbb{R}$, e.g. to be applied separately to each coordinate for multidimensional $\theta$. Denote its values in successive times as $\theta^t$ and $g^t=\partial_{\theta} f(x^t,\theta)|_{\theta=\theta^t}$ as corresponding history of gradients.

Analogously to (\ref{alsr}), to $(\theta^t,g^t)_{t\leq T}$ sequence, we would like to fit parabola $f(\theta)=h+\frac{1}{2}\lambda(\theta-p)^2$, optimizing agreement of derivatives $g^t\approx f'(\theta^t)$ for $f'(\theta)=\lambda(\theta-p)$ using exponentially weakening weights $\eta^{T-t}$:
\be\arg\min_{\lambda,p}\  \sum_{t \leq T} \eta^{T-t} (g^t-\lambda(\theta^t-p))^2\ee
This least squares linear regression leads (\cite{sgd}) to $\lambda$ as $(\theta,g)$ covariance divided by $\theta$ variance, and $\lambda^{-1}$ learning rate gradient descend for averaged position and gradient:
\be \lambda=\frac{\,\overline{g \theta}-\overline{g}\,\overline{\theta}}
{\,\overline{\theta^2}-\overline{\theta}^2 }\qquad \qquad
 p=\overline{\theta}-\lambda^{-1}\,\overline{g}\label{reg}\ee
For $\overline{x}, \overline{g}, \overline{g\theta},\overline{g^2}$ exponential moving averages $(\eta\in(0,1))$:
$$\overline{\theta}^{T+1}=\eta\,\overline{\theta}^T+(1-\eta)\, \theta^T$$
$$\overline{g}^{T+1}=\eta\,\overline{g}^T+(1-\eta)\, g^T$$
$$\overline{g\theta}^{T+1}=\eta\,\overline{g\theta}^T+(1-\eta)\, g^T\, \theta^T$$
$$\overline{\theta^2}^{T+1}=\eta\,\overline{\theta^2}^T+(1-\eta)\, (\theta^T)^2$$

Found $p=\overline{\theta}-\lambda^{-1}\,\overline{g}$ is modeled minimum if $\lambda>0$. Seeing it as gradient descend (using averaged gradient and position), we can e.g. use absolute value and clipping $\epsilon>0$: $$p=\overline{\theta}-\max(|\lambda^{-1}|,\epsilon)\ \overline{g}$$ to handle also negative $\lambda$, and $\lambda\approx 0$ situations (e.g. near inflection point). Finally we can e.g. use $$\theta^{T+1}=\alpha\,p +(1-\alpha)\,\theta^T$$ as parameter evolution step, for some $\alpha\in (0,1]$ describing trust in the parabola model.\\

While online optimization of parameters can be determined by minimization of evolving criterion (\ref{optfun}) like log-likelihood with exponentially weakening weights, there is still required parameter (e.g. $\eta$) of exponential moving average - defining how conservative the model should be. It is usually fixed in data compression (adaptivity) or machined learning (e.g. SGD optimizer) as in a wide range its optimization can only give a relatively tiny improvement.

So in practice such parameter like $\eta$ can be just fixed e.g. optimized over a larger set of data. We could also try to slowly adapt it to local condition if controlling also $\partial\theta^T/\partial \eta$ dependence for used parameters. It would allow to calculate $\partial f(x^T,\theta^T)/\partial \eta$:

$$\frac{\partial f(x^T,\theta^T)}{\partial \eta}(\eta^T)=
\frac{\partial\theta^T}{\partial \eta}(\eta^T) \cdot \nabla_\theta f(x^T,\theta)|_{\theta=\theta^T}$$

Treating them as noisy derivatives again, we can use a gradient methods e.g. after some averaging, or find their linear trend with online linear regression like above.\\

Generally, one could also try to extrapolate e.g. future behavior of parameters based on the recent history, however, it requires extreme caution.

\bibliographystyle{IEEEtran}
\bibliography{cites}
\end{document}